\documentclass[acmsmall]{acmart}

\usepackage{xcolor}
\usepackage{listings}
\usepackage{booktabs}
\usepackage{multirow}
\usepackage{amsmath}
\usepackage{amssymb}
\usepackage{mathpartir}
\usepackage{stmaryrd}
\usepackage{algorithm}
\usepackage{algpseudocode}

\usepackage{mathsprograms}

\usepackage{plantuml}

\definecolor{codegreen}{rgb}{0,0.5,0}
\definecolor{codegray}{rgb}{0.5,0.5,0.5}
\definecolor{codepurple}{rgb}{0.58,0,0.82}
\definecolor{highlightgreen}{rgb}{0.8,1.0,0.8} % Light green
\definecolor{highlightpink}{rgb}{1.0,0.8,0.9}  % Light pink
\definecolor{backcolour}{rgb}{0.95,0.95,0.92}

\newcommand{\llll}{\mathrel{{=}{\llbracket}}}
\newcommand{\rrrr}{\mathrel{{\rrbracket}{\Rightarrow}}}

\lstdefinestyle{solidity_code_style}{
    keywords=[1]{contract, constructor, function, returns, public, private, internal, external, payable, view, pure, override, emit, event, require, if, else, for},
    keywordstyle=[1]\bfseries\color{blue},
    keywords=[2]{uint, uint32, uint256, int, struct, address, bool, string, bytes, mapping},
    keywordstyle=[2]\bfseries\color{codegreen},
    backgroundcolor=\color{backcolour},  
    comment=[l]{//},         % Single-line comments
    morecomment=[s]{/*}{*/}, % Multi-line comments
    commentstyle=\color{codegray},
    numberstyle=\tiny\color{codegray},
    stringstyle=\color{codepurple},
    basicstyle=\ttfamily\tiny,
    breakatwhitespace=false,         
    breaklines=true,                 
    captionpos=b,                    
    keepspaces=true,                 
    numbers=left,                    
    numbersep=5pt,                  
    showspaces=false,                
    showstringspaces=false,
    showtabs=false,                  
    tabsize=2,
    moredelim=[is][\colorbox{highlightgreen}]{@green@}{@endgreen@}, 
    moredelim=[is][\colorbox{highlightpink}]{@pink@}{@endpink@}
}

\usepackage{xcolor}

\AtBeginDocument{%
  }

\setcopyright{acmlicensed}
\copyrightyear{2025}
\acmYear{2025}
\acmDOI{XXXXXXX.XXXXXXX}

\setcitestyle{authoryear}

\newcommand*{\name}{{MPC-EVM}}

\begin{document}

%%
%% The "title" command has an optional parameter,
%% allowing the author to define a "short title" to be used in page headers.
\title{MPC-EVM: Enabling MPC Execution by Smart Contracts In An Asynchronous Manner}

%%
%% The "author" command and its associated commands are used to define
%% the authors and their affiliations.
%% Of note is the shared affiliation of the first two authors, and the
%% "authornote" and "authornotemark" commands
%% used to denote shared contribution to the research.
\author{Yichen Zhou}
\email{yichen.zhou@mail.utoronto.ca}
\affiliation{%
  \institution{University of Toronto}
  \city{Toronto}
  \country{Canada}
}

\author{Chenxing Li}
\affiliation{%
  \institution{Shanghai Tree-Graph Research Institute}
  \city{Shanghai}
  \country{China}}

\author{Fan Long}
\affiliation{%
 \institution{University of Toronto}
 \city{Toronto}
 \country{Canada}}

%%
%% By default, the full list of authors will be used in the page
%% headers. Often, this list is too long, and will overlap
%% other information printed in the page headers. This command allows
%% the author to define a more concise list
%% of authors' names for this purpose.
%% \renewcommand{\shortauthors}{Zhou et al.}

%%
%% The abstract is a short summary of the work to be presented in the
%% article.
\begin{abstract}
  This paper presents MPC-EVM, the first blockchain system that extends the EVM to enable asynchronous MPC invocations by smart contracts during transaction executions without compromising consistency or throughput. MPC-EVM uses an asynchronous execution model to process MPC-invoking transactions in a non-blocking fashion, saving the transaction's progress when it enters an MPC and resuming its execution upon MPC's completion. It also employs an access control mechanism that prevents inconsistent state access and modifications as a result of asynchronous executions. Benchmarking MPC-EVM's throughputs show that the transactions per second (TPS) decreased by less than 3\% compared to the baseline when MPC-invoking transactions are executed alongside regular transactions.
\end{abstract}

\maketitle

\section{Introduction}
Blockchain technology provides a decentralized, tamper‑resistant, and
programmable ledger at global scale. The introduction of smart contracts in
Ethereum extended blockchain functionality beyond simple cryptocurrency
transfers by allowing developers to encode complex transaction logic that is
executed deterministically and enforced by every network node. This trustless
execution paradigm has given rise to a rich ecosystem of decentralized
applications -- from DeFi protocols and supply‑chain tracking to insurance
platforms -- that automate intricate workflows while guaranteeing transparency,
integrity, and resilience without reliance on centralized intermediaries.

However, since every node in a blockchain network must verify every
transaction, all transaction data is inherently public, posing significant
privacy and security challenges. This lack of confidentiality becomes
particularly problematic in critical domains such as finance and identity
management. For instance, decentralized auction contracts on Ethereum publicly
reveal all bidding information, while voting contracts expose individual
voters' choices, thus compromising user privacy. Additionally, the transparent
nature of trading transactions in decentralized exchanges like UniSwap~\cite{uniswap_docs} allows
malicious actors to monitor and analyze incoming user transactions,
facilitating front-running or sandwich attacks that exploit users for profit.

To address these privacy challenges, researchers have employed cryptographic commitments and zero-knowledge proofs (ZKPs)~\cite{ZCash, zksync_protocol}. While effective for basic transactions like private payments, these techniques face fundamental limitations when applied to complex smart contracts. The core constraint stems from ZKP's requirement that all private inputs must originate from a single prover - a condition satisfied in simple cases (e.g., spending UTXOs in Zcash) but problematic for multi-party computations like sealed-bid auctions, where determining the winner requires processing confidential inputs from multiple parties. Previous work such as Hawk\cite{kosba2016hawk} attempted to circumvent this by introducing a trusted coordinator. However, as the coordinator gains unilateral access to all sensitive data while serving as a single point of failure, this approach recreates the very centralized trust models that blockchain technology seeks to eliminate.

Secure multiparty computation (MPC) -- a class of cryptographic protocols that allows multiple participants to jointly evaluate a function on their private inputs without revealing these inputs to any single or small subset of parties~\cite{Cramer_Damgård_Nielsen_2015_Introduction} -- offers a promising solution to blockchain privacy challenges. Many MPC protocols employ secret sharing as a foundational technique: sensitive data (e.g., a numeric value a) is mapped to a random polynomial whose constant term (zero-point evaluation) encodes the original secret. Each participant holds a secret share of the data, corresponding to one point on that polynomial. The randomness in this polynomial ensures that a small subset of shares is uniformly distributed, while the properties of the polynomial enable a sufficiently large set of shares to reconstruct the hidden secret.
MPC computations transform input secret shares into output secret shares by decomposing the target function into a sequence of basic operations (such as multiplication) over secret shares, typically requiring communication rounds among participants.
Similar to blockchain consensus protocols, MPC tolerates a threshold of malicious or faulty parties, ensuring the correctness of outputs and preserving input privacy even in adversarial environments.

A straightforward approach to integrating MPC with blockchain is to execute MPC
computations independently in an off-chain system, which interacts with the
blockchain via standard smart contract interfaces: MPC computation requests are
fetched and initiated from blockchain event logs which smart contracts can
write to, and computation results are subsequently published back to the
blockchain through smart-contract calls. While functional, this approach has
significant drawbacks. 

Firstly, this approach introduces an intermediate "pending" state beyond the EVM's fundamental semantics, where contract calls either succeed or fail immediately. This forces every contract that might directly or indirectly invoke MPC contracts to implement explicit callback interfaces, along with additional hang-up and recovery logic. As a result, it significantly increases the complexity of MPC-involved contracts and disrupts composability between MPC-involved contracts and conventional smart contracts.

Secondly, a smart contract action that involves MPC computations
will be implemented as a \emph{meta-transaction} containing multiple blockchain
transactions and off-chain MPC computation rounds. Because the MPC computations
occur asynchronously relative to blockchain transaction executions, without any
effective synchronization mechanism, other transactions may interleave with the
meta-transaction and produce inconsistent or unintended states. 

Moreover, this
integration weakens blockchain security by relying on the correctness of both
the blockchain consensus and the separate MPC committee. Malicious control
over either the blockchain or MPC component could compromise the integrity of
the ledger state.

\noindent{\textbf{{\name}}.} This paper presents {\name}, the first blockchain platform that natively support MPC computations within smart contracts. {\name} operates with a virtual machine similar to the Ethereum Virtual Machine (EVM), augmented with additional built-in functions that perform private computations, such as secure addition, multiplication, and comparison, using MPC protocols. Developers can write {\name} smart contracts directly in standard Solidity without language modifications. Unlike off-chain MPC approaches, {\name} ensures \emph{sequential consistency} for both regular transactions and MPC-enabled meta-transactions. Specifically, when executing smart-contract operations that involve MPC, {\name} prevents any intermediate computation states from becoming visible or accessible to other transactions.

A key challenge {\name} addresses is achieving high virtual machine performance without sacrificing transaction consistency when MPC-invoking transactions are executed. To mitigate performance impacts when executing high-latency MPC operations, {\name} saves the execution context as part of the blockchain's state, initiates asynchronous MPC computations off-chain, and temporarily suspends the ongoing transaction to allow other unrelated transactions to proceed without delay. As the MPC computation progresses, the MPC module periodically submits special message transactions that update the blockchain state with computation progress and eventually deliver the final MPC results. Once these results become available, the virtual machine retrieves the previously saved execution context and seamlessly resumes transaction execution using the retrieved MPC results. This design enables high-latency MPC computations to run concurrently with other blockchain transactions, effectively preserving high system throughput.

To ensure consistency of MPC meta-transactions relative to other concurrently executing transactions, {\name} employs a novel, \emph{contract-level access control} mechanism. When {\name} saves the execution context of an MPC-invoking transaction, it simultaneously locks the initiating contract, preventing other transactions from accessing or modifying its state during the ongoing MPC computation. Furthermore, {\name} enforces an additional constraint that the resumed execution of MPC-invoking transactions cannot access external contract states. Together, these mechanisms preserve sequential consistency while enabling parallel execution of MPC computations alongside blockchain transactions.

\noindent{\textbf{Experimental Results.}} We implemented a prototype of {\name} and developed 5 benchmark smart contracts involving MPC computations, including privacy-preserving voting and auction scenarios. Our experimental evaluation demonstrates that {\name} successfully executes all benchmark contracts as expected. Moreover, the results confirm that high-latency MPC operations effectively run in parallel with regular blockchain transactions. When executed alongside MPC-enabled transactions, the throughput for regular transactions experienced only a modest decrease of approximately 2.6\% on average, confirming {\name}’s efficiency and practicality.

\noindent\textbf{Contributions.} This paper makes the following contributions:
\begin{itemize}
  \item \textbf{{\name}:} We introduce {\name}, the first blockchain platform to natively support MPC computations within smart contracts.

  \item \textbf{Virtual Machine Design:} We propose a novel virtual machine architecture that augments the standard EVM with MPC-specific builtin functions, enabling asynchronous execution of MPCs and MPC-invoking transactions to achieve high transaction throughput.

  \item \textbf{Contract-Level Access Control Policy:} We develop a novel access control mechanism at the granularity of individual contracts. When integrated with our virtual machine design, this mechanism ensures sequential consistency of both regular and MPC-enabled smart contract transactions.
\end{itemize}

The rest of the paper is organized as follows. Section ~\ref{sec:background} presents background for blockchain, smart contracts, and MPC. Section~\ref{sec:motivating_example} presents a motivating example for {\name}. Section~\ref{sec:formal_design} and Section~\ref{sec:implementation} presents the design and the implementation of {\name}. We evaluate {\name} in Section~\ref{sec:evaluation}. We discuss related work in Section~\ref{sec:related_work} and finally conclude in Section~\ref{sec:conclusion}.
\section{Background} \label{sec:background}
\textbf{Blockchain}: Blockchains are decentralized ledgers where a network of servers called nodes processes user-submitted transactions and updates the ledger's state. Each node maintains a consistent view of the ledger state by locally executing each transaction and relies on a distributed consensus protocol to ensure that the transactions are executed by all nodes in the same order. The transactions are processed in batches called blocks, and each processed block is linked to a previous one via a cryptographic hash, forming a chain that prevents the retroactive alteration of transaction history. The Ethereum blockchain, whose native cryptocurrency is Ether (ETH), operates as a distributed state machine whose state is updated by transaction executions in the Ethereum Virtual Machine (EVM). A block of transactions is executed by the EVM to generate the pending modifications to the blockchain state. When the block is approved by the consensus process, the transactions are committed, and the pending state changes are written into each node's persistent storage\\
\textbf{EVM}: EVM is a stack-based, Turing complete virtual machine that executes bytecodes instructions composed from a predefined list of EVM opcodes. Opcodes instruct the EVM to perform various low-level operations, such as popping the previous instruction's result from the stack or reading a value from storage. Transactions invoke EVM's execution of smart contract bytecodes as described later. Important guarantees provided by the EVM include transaction atomicity and consistency. EVM uses a sequential execution model where each transaction is executed to completion before the next one gets processed. Additionally, if a transaction execution fails for various reasons, EVM discards the pending state changes made by this transaction and reverts back to the most recent valid state. \\
\textbf{Smart Contracts}: Smart contracts are programs developed by users to define customized transaction rules on the blockchain. They are compiled into EVM bytecodes when deployed onto the blockchain in a create transaction, and they contain functions that can be called by subsequent transactions. Calling smart contract functions execute the deployed bytecode and modify the blockchain state according to the smart contract's logic in a deterministic manner. \\
\textbf{Accounts}: Ethereum accounts can be either an externally owned account (EOA) controlled by a user via a private key or a contract account hosting a smart contract. Each account on Ethereum has a unique address, an associated balance, and nonce, which for an EOA is a counter for the number of transactions sent by this account. Contract accounts have an additional code and state variable storage. EOAs can send transactions to transfer balances to another EOA or execute a smart contract function. \\
\textbf{Secure multiparty computation (MPC)}: MPC is a family of cryptographic protocols in which \(n\) mutually distrustful parties collaboratively evaluate a function \(f(x_1, x_2, ..., x_n) = (y_1, y_2, ..., y_m)\), where \(x_i\) is party \(i\)'s secret input and
\(y_j\) is the \(j^{\text{th}}\) output, in a way such that \((y_1, y_2, ... y_m)\) can be learned by all parties while each party's input remains secret as long as the security assumptions of the protocol is met. Specifically, actively secure MPC protocols, which guarantee output correctness while maintaining input privacy against malicious parties that could collaborate to sabotage the protocol in an attempt to deduce the honest parties' inputs, require that no more than \(t\) out of \(n\) parties are malicious, where \(t\ < \frac{n}{3}\).

One foundational technique for implementing such protocols is Shamir-Secret Sharing (SSS). In SSS-based MPC protocols, a secret value $ a $ is encoded into a random polynomial $ f_a(x) $ of degree $ t $, where the constant term $ f_a(0) = a $ represents the secret itself. Each party receives a share $ [a]_t^{(i)} = f_a(i) $, corresponding to a point on the polynomial. And the notation $ [a]_t$ represents the collective secret shares held by all participants. The randomness of the polynomial ensures that any subset of $ t $ or fewer shares reveals no information about $ a $, while any $ t+1 $ or more shares can reconstruct the secret through polynomial interpolation. 

Building on secret sharing, MPC protocols enable parties to perform computations on shared data without revealing the underlying secrets. Addition between two secret-shared values is particularly straightforward due to the linearity of polynomial sharing. Given shares $ [a]_t $ and $ [b]_t $, each party can locally compute $ [a+b]_t $ by simply adding their respective shares $ [a + b]_t^{(i)} = [a]_t^{(i)} + [b]_t^{(i)} $, corresponding to the evaluation of the polynomial $ f_{a+b}(x) = f_a(x) + f_b(x) $. 

Multiplication between two secrets-shared values, however, introduces complexity due to its non-linear nature. When parties multiply their local shares, the result $ [ab]_{2t}^{(i)} = f_a(i) \cdot f_b(i) $ corresponds to a polynomial of degree $ 2t $. This necessitates the \emph{re-share} process to reduce the polynomial degree back to $ t $. In this process, each party $ i $ distributes a new secret share $ [h_i]_t $ of their local product $ h_i = [ab]_{2t}^{(i)} $. These secret shares can be linearly combined with Lagrange coefficients to obtain $ [ab]_t $ via polynomial interpolation. This process involves significant communication overhead, requiring 2 rounds of all-to-all communication in the semi-honest security model and many more in the malicious security model.

To defend against malicious adversaries who may deviate from the protocol, MPC protocols incorporate mechanisms for verifying correctness and enforcing consistency. One key technique is \textit{verifiable secret sharing (VSS)}, which binds parties to their shares via cryptographic commitments. A \emph{commitment scheme} 
$C = \text{Commit}(m, r)$
takes a message $m$ and random input $r$, producing a commitment $C$ that is both \textit{hiding} (it reveals no information about $m$) and \textit{binding} (it is infeasible to find $m' \neq m$ with $\text{Commit}(m', r') = C$). 

In VSS, a dealer shares a secret \(a\) by choosing a random value \(r\) and distributing \([a]_t^{(i)}\) and \([r]_t^{(i)}\) to each party \(i\). Along with these shares, the dealer provides a public commitment \(C_i = \text{Commit}([a]_t^{(i)}, [r]_t^{(i)})\). The commitment scheme is homomorphic, allowing \( \text{Commit}(m_1 + m_2,\, r_1 + r_2) \) to be computed directly from \( \text{Commit}(m_1,\, r_1) \) and \( \text{Commit}(m_2,\, r_2) \). Using Lagrange interpolation on the commitments \( C_i \), parties can derive \( \text{Commit}(a,\, r) \) from the linear combination of public commitments \(C_i\). Intuitively, the binding property ensures each commitment corresponds uniquely to its secret data or secret share, guarantees \textit{integrity}, while the hiding property preserves the \emph{privacy} of the secret. Once \(t+1\) or more honest parties confirmed that their \([a]_t^{(i)}\) and \([r]_t^{(i)}\) are consistent with the corresponding commitments \(C_i\), they collectively possess enough correct information to reconstruct \(a\), guaranteeing availability of the secret.

If any share is found to be inconsistent with its public commitment, a dispute resolution is triggered: the accused party must reveal the inputs of its commitment to prove the validity of its share. Failing this or having invalid shares results in exclusion. The protocol then discards incorrect shares and uses valid shares for polynomial interpolation, ensuring the secret can still be recovered even in the presence of corrupted participants.
\section{Motivating Example} \label{sec:motivating_example}

We now present a motivating example to demonstrate how {\name} simplifies the development and execution of decentralized voting contracts. Such voting contracts play a central role in the governance mechanisms of Decentralized Autonomous Organizations (DAOs), including MakerDAO~\cite{makerdao} and Aragon~\cite{aragonDAO}. Traditional blockchain-based voting implementations require voters to submit votes as transactions, inevitably exposing individual choices publicly. In contrast, our example leverages MPC to ensure voter privacy by encapsulating votes as secret MPC inputs, revealing only the final outcome — the proposal receiving the highest vote count — as the public result.

\begin{lstlisting}[float=t, label={lst:mpcvote}, caption={Simplified Code Snippet of MPC Voting Contract}]
contract MPCVote {
    address organizer; //deployer of this contract
    string[2] proposals; //name of the proposals being voted on
    address[] voters; //addresses of the voters
    uint[] weights; //weight of each voter's vote
    ...
    uint public startableTime; //earliest time the mpcVote() function can be called
    MPCTxMgr mpcMgr;
    uint public minDeposit; //minimum deposit required for each voter
    uint cid;
    mapping(address => uint) deposits;

    constructor(address[] voters_, string p0, string p1, uint _minDeposit, uint voteCircuitId) {
        //initialize mpgMgr with builtin constant address
        mpcMgr = MPCTxMgr(MPCTXMGR_ADDR);
        organizer = msg.sender;
        minDeposit = _minDeposit;
        voters = voters_;
        proposals[0] = p0;
        proposals[1] = p1;
        startableTime = block.timestamp + 3600; //give voters 1 hour to make the required deposit
        cid = voteCircuitId;
    }

    function deposit() external payable { ... }
    function withdraw(uint amount) external validAmount(amount) { ... }

    function mpcVote() external isOrganizer validStartTime {
        //voters who did not make the required amount of deposit have zero weight 
        for (uint i = 0; i < voters.length; ++i) {
            if(deposits[voters[i]] < min_dep) {
                weights[i] = 0;
            } else { weights[i] = deposits[voters[i]] }
        }
        //invoke MPC vote procedure
        uint[] memory results = mpcMgr.enter_mpc(cid, weights);
        //process result
        processVoteResult(results);
    }

    function processVoteResult(uint[] results) internal {
        if (results[results.length-2] != 0) { //cheater caught
            address cheater = voters[results.length-1];
            processCheater(cheater);
        }
        else { //MPC completed successfully
            winnerId = results[0];
            succeeded = true;
            ...
        }
    }

    function processCheater(address cheater) internal {
        //distribute the cheater's deposits to other voters and organizer
        ...
    }
    ...
}
\end{lstlisting}

\begin{lstlisting}[float=t, label={lst:votecircuit}, caption={Pseudocode for constructing a n-party voting for 2 proposals}]
/* x[i][j] where 0 <= i < n and j- <= j <= 1 is a secret shared 0 or 1, indicating whether voter i voted for the jth proposal. 
   w is the publicly known list of voter weights */
fn 2_proposal_voting(x[n; 2]: Secret, w[n; 1]: Public) -> winner {
    s1 = new Secret[n]
    s2 = new Secret[n]
    s1[0] = MPCMultByConst(x[0][0], w[0])
    s2[0] = MPCMultByConst(x[0][1], w[0])
    for i in 1 .. n:
        s1[i] = MPCAdd(s1[i-1], MPCMultByConst(x[i][0], w[i]))
        s2[i] = MPCAdd(s2[i-1], MPCMultByConst(x[i][1], w[i]))
	
    //Compare the vote tallys
    let [max], [max_id] = MPCCompare(s1[n-1], 0, s2[n-1], 1)
	
    winner = max_id
}
\end{lstlisting}

\noindent \textbf{Voting Contract.} Listing~\ref{lst:mpcvote} shows a simplified snippet of the \texttt{MPCVote} smart contract implemented in Solidity, demonstrating secure voting through MPC using {\name}. Non-essential details unrelated to MPC computation are omitted for clarity. {\name} includes a built-in contract called \emph{MPC Transaction Manager}, available at a predefined constant address within the EVM address space, offering a suite of library functions to facilitate MPC computations (line 15 in Listing~\ref{lst:mpcvote}). Additionally, {\name} offers basic MPC computation primitives—including addition, multiplication, multiplication by a constant, and comparison—that developers can leverage to construct custom MPC circuits. These circuits process both secret and public inputs and produce secure computation results.

\noindent \textbf{MPC Computation.} Listing~\ref{lst:votecircuit} shows simplified pseudocode illustrating the MPC computation logic for our voting example. It leverages several MPC primitives provided by {\name}, including \texttt{MPCMultByConstant()}, \texttt{MPCAdd()}, and \texttt{MPCCompare()}. The computation receives secret input arrays from $n$ different voters (denoted by \texttt{x} at line 3 in Listing~\ref{lst:votecircuit}), with each element indicating voter support (1) or non-support (0) for a given proposal. Additionally, a public input array specifies each voter's weight. The computation proceeds by calculating the weighted sum of votes for each proposal using two secret arrays, \texttt{s1} and \texttt{s2} (lines 4--10). Finally, it publicly reveals the winning proposal by comparing these two totals (line 13).

\noindent \textbf{Usage Scenario.} To initiate the voting process, an organizer first submits a transaction to the built-in MPC transaction manager contract to register the MPC computation logic shown in Listing~\ref{lst:votecircuit}, obtaining a unique \texttt{circuitId}. Next, the organizer deploys the voting contract by invoking its constructor (line 13 in Listing~\ref{lst:mpcvote}). The constructor call provides the obtained \texttt{circuitId} (\texttt{voteCircuitId} at line 22), along with voter addresses (line 18), the minimum deposit required for voting eligibility (line 17), and the two candidate proposals (lines 19--20). While this example illustrates voting between two proposals for simplicity, the approach easily generalizes to handle an arbitrary number of proposals. Additionally, the constructor enforces a minimum waiting period—set to one hour after contract deployment (line 21)—allowing voters sufficient time to deposit tokens and prepare their votes.

To participate in voting, voters invoke the \texttt{deposit()} function (line 25), depositing native tokens into the contract. These deposits serve a dual purpose: determining voting weights and acting as stakes that can be slashed in response to malicious behaviors during MPC computations. After voting concludes, voters can reclaim their deposits by invoking the \texttt{withdraw()} function (line 26). For simplicity, we omit detailed implementations of these two functions; in practice, they update the internal \texttt{deposits} mapping accordingly upon processing deposits or withdrawals.

Once the designated waiting period has elapsed, the organizer invokes the \texttt{mpcVote()} function (line 28) to tally the votes. This function first verifies whether each voter has met the required minimum deposit (lines 29--34). It then initiates the MPC-based vote-counting process by calling the \texttt{enter\_mpc()} function of the built-in MPC manager contract (line 36). The \texttt{enter\_mpc()} function accepts two parameters: the circuit identifier (\texttt{cid}), specifying the registered MPC computation logic, and the voter weight array, providing the public inputs for the MPC circuit. Finally, \texttt{mpcVote()} processes the MPC results by invoking the \texttt{processVoteResult()} function.

The MPC computation result in our example is returned as an array containing three elements. The first element represents the computed outcome—the identifier of the winning proposal. The second element is a status flag indicating whether the MPC execution completed successfully or failed due to malicious participant behavior. The \texttt{processVoteResult()} function first examines this status flag. If the flag indicates failure (a non-zero value), the function identifies and penalizes the malicious participant (lines 42--45). Otherwise, it extracts the winning proposal's identifier and concludes successfully (lines 47--48).

We now describe how {\name} processes transactions involving MPC computations in this example. Specifically, when the organizer submits a transaction invoking the \texttt{mpcVote()} function, {\name} executes the following steps:

\noindent \textbf{Save Execution State before MPC Computation.} 
The {\name} virtual machine executes transactions normally—just like standard EVM transactions—up to the point where an MPC computation is invoked (line 38 in Listing~\ref{lst:mpcvote}). At this point, {\name} saves the current execution state into a dedicated region of the blockchain state. This saved state includes essential data such as the EVM stack, the program counter, and any uncommitted blockchain state changes. After saving this context, {\name} initiates the MPC computation asynchronously and suspends the transaction execution, marking it with a special ``paused'' status to indicate that execution should be resumed once the MPC computation completes.

\noindent \textbf{Parallel Execution with Access Policy.}
When the execution of the \texttt{mpcVote()} transaction is paused to await the completion of the MPC computation, {\name} continues processing other transactions concurrently to maintain high throughput. However, {\name} places a lock on the \texttt{MPCVote} contract, preventing execution of any transaction that attempts to access or modify its state. For instance, new transactions calling \texttt{deposit()} or \texttt{withdraw()} on \texttt{MPCVote} will be deferred due to their dependency on the locked contract state. 
Since \texttt{mpcVote()} does not access state outside of the \texttt{MPCVote} contract, this locking mechanism ensures that {\name} only processes transactions without dependencies on paused transactions, thus guaranteeing the consistency of transaction execution outcomes.

\noindent \textbf{Resuming the \texttt{mpcVote()} Transaction.}
Once the MPC computation completes, the MPC module of {\name} generates a special message transaction containing the MPC computation results (i.e., \texttt{results} at line 36 in Listing~\ref{lst:mpcvote}). Upon processing this message transaction, {\name} restores the previously saved execution state, removes the lock from the \texttt{MPCVote} contract, and resumes the paused execution at line 38 in Listing~\ref{lst:mpcvote} using the returned MPC results. Since there are no subsequent MPC computations in this example, the transaction execution continues uninterrupted and successfully completes as a standard EVM transaction.

\section{MPC-EVM Design} \label{sec:formal_design}
This section describes the design of {\name}. We formally explain how {\name} extends the Ethereum Virtual Machine (EVM) by incorporating MPC computation capabilities and present how it supports parallel execution of regular transactions and MPC computations without sacrificing transaction consistency, enforced via our access control policy. Similar to EVM, the architecture of {\name} consists of two layers: a virtual machine execution layer that governs individual transaction execution, and a block processing layer responsible for updating blockchain state across multiple transactions within a block.

\begin{figure}[h]
    \begin{flalign*}
        &\langle prog \rangle ::= \text{program} \ \langle instr \rangle^* \\
        &\langle instr \rangle ::= \langle EVMinstr \rangle \mid \langle MPCinstr \rangle \\
        &\langle MPCinstr \rangle ::= y = \texttt{enter}\_\texttt{mpc}(\textsf{cid}, x) \\
        &\langle EVMinstr \rangle ::= \texttt{call}(\textsf{ad}, \text{m}) \mid \ldots
    \end{flalign*}
    \setlength{\abovecaptionskip}{0pt}
    \caption{Core MPC-EVM language syntax rules. \(x^*\) represents zero or more occurrences of \(x\)}
    \label{fig:mpcevm-lang}
    \setlength{\belowcaptionskip}{0pt}
\end{figure}

\subsection{{\name} Language}
Figure~\ref{fig:mpcevm-lang} shows the syntax of a simplified programming language used to illustrate our design. The language extends standard EVM instructions with an additional instruction, \texttt{enter\_mpc}, to explicitly invoke MPC computations. Specifically, the \texttt{enter\_mpc} instruction triggers an MPC computation identified by \textsf{cid}, using variable \texttt{x} as the public input. The result of this MPC computation is then stored in a local variable \texttt{y}. Additionally, the language includes a simplified version of EVM's \texttt{call} instruction, which transfers execution to a designated method \texttt{m} within a smart contract at the specified address \textsf{ad}.

Note that our simplified language uses local variables, whereas the standard EVM operates exclusively with an execution stack; this difference is introduced solely for the simplification of the semantics presentation. Additionally, we simplified the \texttt{call} instruction and omitted other EVM instructions that are not essential to understanding the core design of {\name}. Our prototype implementation fully supports all standard EVM instructions and provides additional interfaces for managing and registering new MPC computations. See Section~\ref{sec:implementation} for further details on these implementation features.

\begin{figure}
\begin{mathpar}
\inferrule{
    y = \texttt{enter\_mpc}\ (\textsf{cid}, x) \in\textsf{inst}(\lsconf(\pcconf))\\
    v := \lsconf(x)\\
    \calladdrs = \{\addr\} \\
    M := \mathsf{initMPC}(\textsf{cid}, v) \\
    \mpcstates' := \mpcstates[\addr\mapsto (y, M, (\ameth, \addr, \lsconf) \circ \rtaskconf)] 
}{
    (\_, (\ameth, \addr, \lsconf) \circ \rtaskconf,
    \mpcstates, \calladdrs)\llll y = \texttt{enter\_mpc}\ (\textsf{id}, x)\rrrr (\_, \epsilon, \mpcstates', \calladdrs) 
}\and

\inferrule{
    \texttt{call}\ (\addr',\ameth') \in\textsf{inst}(\lsconf(\pcconf))\\
    \calladdrs' := \calladdrs \cup \{\addr'\} \\
    \lsconf' := \mathsf{locInit}(\gsconf, \ameth')
}{
    (\gsconf, (\ameth, \addr, \lsconf) \circ \rtaskconf, \_, \calladdrs)\llll\texttt{call}\ (\addr', \ameth') \rrrr (\gsconf, (\ameth', \addr', \lsconf') \circ (\ameth, \addr, \lsconf) \circ \rtaskconf, \_, \calladdrs') 
}\and
\end{mathpar}
\caption{Program Semantics in {\name}. 
For a function $f$, we use $f[a\mapsto b]$ to denote a function $g$ such that $g(c)=f(c)$ for all $c\neq a$ and $g(a)=b$. The function $\instrOf$ returns the instruction at some given control location while $\mathsf{next}$ gives the next instruction to execute. We use $\circ$ to denote sequence concatenation. 
$\mathsf{locInit}(\bsconf,\ameth)$ represents the initial state of $\ameth$.
We use $\epsilon$ in the execution stack to represent the end of an execution.
The function $\textsf{initMPC}$ starts the MPC computation rounds and returns the initial MPC state.}
\label{fig:mpcevm-op}
\end{figure}

\subsection{Operational Semantics}

A program configuration in {\name} is represented as a tuple $\mu = (\gsconf, \rtaskconf, \mpcstates, \calladdrs)$, where $\gsconf$ denotes the persistent blockchain state, $\rtaskconf$ denotes the current call stack, $\mpcstates$ is a mapping from each contract address to its saved execution state for ongoing MPC transactions, and $\calladdrs$ is the set of contract addresses accessed by the transaction execution thus far.

Call frames within the $\rtaskconf$ are represented as tuples $(\ameth, \addr, \lsconf)$, where $\ameth \in \mathbb{M}$ denotes the name of the invoked function, $\addr \in \mathbb{A}$ denotes the address of the contract containing the function $\ameth$, and $\lsconf$ represents the local variable valuation environment, including the program counter $\pcconf$. Each address $\addr$ in $\mpcstates$ maps to a tuple $(y, M, \rtaskconf)$, where $y$ is the local variable designated to store MPC computation results, $M$ holds the state required by the MPC modules, and $\rtaskconf$ captures the saved execution state.

Figure~\ref{fig:mpcevm-op} presents the small step semantics for \texttt{enter\_mpc} and \texttt{call} instructions in {\name}. The notation $\mu \llll a \rrrr \mu'$ represents the state transition of $\mu$ to $\mu'$ after execution the operation $a$. For \texttt{enter\_mpc}, the transition initiates the MPC computation process via the function \textsf{initMPC}, saves the current execution state into the mapping $\mpcstates$, and then quit the current execution by setting the stack to $\epsilon$. Note that the transition also checks $\calladdrs = \{\addr\}$ to ensure the execution has never accessed states of other contracts before.
The transition for \texttt{call} appends a new call frame on top of the execution stack. It also updates $\calladdrs$ to mark the execution accessed a new contract at the address $\addr'$.

The execution of an externally invoked method $\ameth$ within a contract located at address $\addr$ is represented as a sequence of transitions $\rho = \mu_0 \llll a_1 \rrrr \mu_1 \llll a_2 \rrrr \ldots \llll a_n \rrrr \mu$, beginning with the initial configuration $\mu_0 = (\gsconf, (\ameth, \addr, \lsconf_0), \mpcstates, \{\addr\})$, where $\lsconf_0 = \mathsf{locInit}(\gsconf, \ameth)$ defines the initial local state of the function $\ameth$. This sequence of transitions results in the final configuration $\mu = (\gsconf, \epsilon, \mpcstates', \calladdrs)$, where the execution stack is empty. We denote this transition sequence as $\mu = \rho_m(\mu_0)$, indicating that execution $\rho$ of method $\ameth$ transforms the initial state $\mu_0$ into the final state $\mu$.

For brevity, our semantics omit the logic related to gas calculation and enforcement. In standard EVM execution, each execution state also maintains a gas counter tracking resource usage, and transactions terminate prematurely with an error if gas consumption exceeds the specified limit. Our implementation of {\name}'s virtual machine adheres to similar gas-handling rules and additionally allows for defining a custom gas cost for the \texttt{enter\_mpc} instruction.

\subsection{Transaction Execution}

\begin{figure}
\begin{mathpar}
\inferrule{
    (\nonce_s,\bbbalance_s,\textsf{act}_s,\progcode_s) = \bsconf(\addr_s) \\
    \bbbalance_s \geq \bbbalance_0 \\ 
    \addr_n \in \mathbb{A}\ fresh \\ 
    \gsconf_0 := \mathsf{updBalNonceAndInit}(\bsconf,\addr_s, \addr_n,\bbbalance_0, c) \\
    \lsconf_0 := \mathsf{locInit}(\gsconf_0,\ameth) \\
    (\gsconf_1, \epsilon, \mpcstates',\calladdrs) := \rho_{\ameth}(\gsconf_0, (\ameth, \addr_n, \lsconf_0), \mpcstates, \{\addr_n\}) \\
    \mpcstates \cap \calladdrs = \emptyset \\
}{
    (\height, \bsconf, \mpcstates) \llll \texttt{createTx}\ (\addr_s,\nonce_s,\ameth,\bbbalance_0,\progcode) \rrrr^{t} (\height, \bsconf_1, \mpcstates')
} \and
\inferrule{
    (\nonce_s,\bbbalance_s,\textsf{act}_s,\progcode_s) = \bsconf(\addr_s) \\
    \bbbalance_s \geq \msgvalue \\
    \bsconf_0 := \mathsf{updBalNonce}(\bsconf,\addr_s, \addr_t, \msgvalue) \\
    \lsconf_0 := \mathsf{locInit}(\gsconf_0,\ameth) \\
    (\gsconf_1, \epsilon, \mpcstates',\calladdrs) := \rho_{\ameth}(\bsconf_0, (\ameth, \addr_t, \lsconf_0), \mpcstates,\{\addr_t\}) \\ 
    \mpcstates \cap \calladdrs = \emptyset \\
}{
    (\height, \bsconf, \mpcstates) \llll \texttt{regularTx}\ (\addr_s,\nonce_s,\addr_t,\ameth,\msgvalue) \rrrr^{t} (\height, \bsconf_1, \mpcstates')
} \and
\inferrule{
    (\nonce_s,\bbbalance_s,\textsf{act}_s,\progcode_s) = \bsconf(\addr_s) \\
    (y, M, \rtaskconf) = \mpcstates(\addr_m) \\
    M' := \mathsf{updMPC}(M, \addr_s, \addr_m, \textsf{msg}) \\
    \mpcstates'=\mpcstates[\addr_m\mapsto (y, M', \rtaskconf)] \\
    \bsconf_1 := \mathsf{updNonce}(\bsconf,\addr_s) \\
}{
    (\height, \bsconf, \mpcstates) \llll \texttt{mpcmessageTx}\ (\addr_s,\nonce_s,\addr_m,\textsf{msg}) \rrrr^{t} (\height, \bsconf_1, \mpcstates')
} \and
\inferrule{
    (\nonce_s,\bbbalance_s,\textsf{act}_s,\progcode_s) = \bsconf(\addr_s) \\
    (y, M, (\ameth, \addr, \lsconf) \circ \rtaskconf) = \mpcstates(\addr_m) \\
    \lsconf_0 = \lsconf[\pcconf\mapsto \mathsf{next}(\lsconf(\pcconf))][y\mapsto \textsf{ret}] \\
    \mpcstates_0 = \mpcstates \setminus \addr_m\\
    \bsconf_0 := \mathsf{updNonce}(\bsconf,\addr_s) \\
    (\gsconf_1, \epsilon, \mpcstates',\calladdrs) := \rho_{\ameth}(\bsconf_0, (\ameth, \addr, \lsconf_0) \circ \rtaskconf, \mpcstates_0,\{\addr\}) \\
     \mpcstates_0 \cap \calladdrs = \emptyset \\
}{
    (\height, \bsconf, \mpcstates) \llll \texttt{mpcretTx}\ (\addr_s,\nonce_s,\addr_m,\textsf{ret}) \rrrr^{t} (\height, \bsconf_1, \mpcstates')
} \and
\end{mathpar}
\caption{Transactions execution rules. 
\textsf{updNonce()} updates the blockchain state by increment the nonce by one.
\textsf{updBalNonce()} in addition updates the blockchain state balances following a simple transfer.
\textsf{updBalNonceAndInit()} further initializes the blockchain state for a new deployed contracts.
$\textsf{updMPC}(M, \addr_s, \addr_m, \textsf{msg})$ denotes the MPC procedure of processing the received message $\textsf{msg}$. 
}
\label{fig:mpcevm-tx}
\end{figure}

A blockchain state in {\name} is represented by the tuple $\sigma = (h, \bsconf, \mpcstates)$, where $h$ denotes the current block height, $\bsconf$ denotes the persistent blockchain state containing all account information, and $\mpcstates$ maintains the saved execution states of ongoing MPC transactions. Specifically, $\bsconf$ maps each address $\addr$ to $\bsconf(\addr) = (\nonce, \bbbalance, \textsf{vars}, c)$, where $\nonce$ is the account nonce, $\bbbalance$ is the account balance, $\textsf{vars}$ represents the persistent storage variables of a contract, and $c$ is the immutable contract code.

Figure~\ref{fig:mpcevm-tx} presents the transaction execution rules in {\name}. We denote the blockchain state transition from state $\sigma$ to state $\sigma'$ after executing transaction $\textsf{tx}$ as $\sigma \llll \textsf{tx} \rrrr^{t} \sigma'$. {\name} supports four types of transactions: (1) create transactions, which deploy new contract accounts and initialize their balances; (2) normal transactions, which invoke smart contract methods and transfer balances; (3) MPC message transactions, used for inter-party communication within MPC committees; and (4) MPC result transactions, which indicate the completion of MPC computations. The first two transaction types exist in standard EVM, while the latter two are specifically introduced by {\name} to facilitate MPC integration.

The transition labeled \texttt{createTx} deploys a new contract at a fresh address $\addr_n$, initializes the contract state, and updates both the balance and the nonce accordingly. Subsequently, it starts executing the specified constructor method $\ameth$. Similarly, the \texttt{regularTx} transition initiates execution of the invoked method $\ameth$ in an existing contract.

Note that both transitions enforce the condition $\mpcstates \cap \calladdrs = \emptyset$ to maintain our access control policy. Specifically, this condition ensures that a transition does not access contracts involved in ongoing MPC transactions that are currently suspended and awaiting MPC results. Executing transactions that interact with these locked contracts could result in inconsistent accesses and modifications, thereby violating the consistency guarantees provided by {\name}.

The transition labeled \texttt{mpcmessageTx} represents scenarios where the asynchronous MPC module in {\name} receives intermediate broadcast messages from other MPC committee members. At this stage, the MPC module is not yet ready to produce the final MPC results but the broadcast may trigger additional MPC execution logic to update its internal MPC state. This state-update logic is captured by the \textsf{updMPC()} function, which may trigger additional MPC messages since many MPC operations require multiple communication rounds with inter-dependencies. {\name} leverages blockchain transactions to realize MPC broadcast communications, enhancing robustness.

The transition labeled \texttt{mpcretTx} corresponds to the scenario in which the MPC module returns the final computation result, denoted by $\textsf{ret}$. In this transition, the previously saved execution state is retrieved from $\mpcstates(\addr_m)$, the returned result is assigned to the local variable $y$, and the program counter is advanced to the next instruction. Execution then resumes from this restored state. To enforce our access control policy, the transition also verifies the condition $\mpcstates \cap \calladdrs = \emptyset$, ensuring that the resumed execution does not interact with other MPC transactions that remain suspended.

\section{Implementation} \label{sec:implementation}
We implemented MPC-EVM on top of AptosEVM \cite{zero_gravity_core}, a hybrid blockchain system that integrates the Conflux EVM \cite{confluxrust} with the Aptos blockchain \cite{aptoscore}. All existing EVM opcodes are supported by this system. We built MPC-EVM from AptosEVM by making three main additions: an MPC execution module that performs the invoked MPCs, a modified virtual machine that realizes the asynchronous execution of MPC transactions, and an MPC transaction manager smart contract that manages the MPC states and serves as an interface between on-chain transaction executions and off-chain MPC computation.

\subsection{System Architecture}

At a high level, a typical blockchain system consists of the following core modules:  
\begin{enumerate}
    \item \textbf{P2P Network Module}: Manages node discovery, peer connections, and message propagation across the network. It adheres to P2P protocols to broadcast transactions, blocks, and other network messages while coordinating communication between nodes.  
    \item  \textbf{Consensus Module}: Operates atop the P2P network to exchange blocks and transactions with peers. Through the consensus protocol, it ensures blockchain progression, maintains transaction pools, generates new blocks, submits blocks/transactions to the execution module, and records the execution results.
    \item \textbf{Execution Module}: Acts as a stateless executor that processes blocks passed down from the consensus module. It initiates transaction execution from specified ledger state versions and interacts with the storage module to access ledger states.  
    \item \textbf{Storage Module}: Maintains multi-versioned ledger states and provides read/write interfaces for historical and current blockchain data.  
\end{enumerate}

The architecture is enhanced with an additional \textbf{MPC execution module}, designed to integrate MPC computation with existing blockchain operations. When the transaction execution module initiates an MPC execution request, the execution engine temporarily pauses the relevant transaction, saves its contextual state, and proceeds with subsequent transactions without waiting for the MPC transaction to complete. Upon completing the current execution task, the execution module propagates MPC request metadata along with other execution results to the consensus module. The consensus module notifies the MPC execution module of an MPC execution request only after the block containing the request has achieved consensus finality.

During MPC execution processing, the module utilizes the network module for direct P2P communication, while relying on the consensus module's broadcast mechanism for messages requiring universal acknowledgment from all honest participants. These broadcast-dependent messages are encoded as specialized transactions, undergo standard consensus finalization, and are then routed back to the MPC execution module.

Upon completing MPC operations, the module generates a designated transaction type containing execution resumption instructions. When encountering this transaction, the execution module retrieves the preserved contextual state and seamlessly resumes processing the suspended transaction execution.

\subsection{MPC Execution Module}
The MPC execution module implements the information-theoretically secure MPC protocol proposed by \cite{Cramer_Damgård_Nielsen_2015_IT_Protocol}, which is Shamir-Secret-Sharing (SSS) based and robust against t < \(\frac{n}{3}\) malicious MPC parties. %We assume that this honest majority assumption holds for any MPC invoked by transactions on MPC-EVM. 
Although other protocols that offer lower communication without security compromise do exist, we selected this classic protocol for implementation to focus on integrating MPC with the blockchain system rather than delving into the intricate engineering details of MPC itself.

% Additionally, our implementation differs from the original protocol in that it immediately aborts with a malicious party's id when it is caught.

% In the MPC execution module, a \textit{manager} task is spawned when the node is bootstraped. It runs perpetually in parallel with the other blockchain modules and listens for MPC execution requests passed down from the consensus layer. 

% The MPC execution module will receive the following two types of items from the consensus module:
% \begin{enumerate} 
%     \item a transaction invoking MPC for the first time 
%     \item a broadcast MPC message, specifically a result attestation message, that enabled an MPC transaction to resume its execution, which ran into another MPC invocation    
% \end{enumerate}  

% Comparison between two secret-shared values is a much more complex MPC operation that requires a circuit consisting of hundreds of addition and multiplication gates, and it is implemented using the protocol proposed by. 

% The fundamental building block of an arithmetic circuit is a gate that performs a single arithmetic operation, including secret addition and multiplication. The arithmetic circuit of MPC voting and auctions, whose logic is in essence a series of comparisons between secret shared values to identify the maximum, are built using a composition of the MPC comparison modules.

When an MPC execution request is received, this module spawns a new \emph{executor} that loads the requested arithmetic circuit, which is natively implemented in the MPC execution module, and begins its evaluation. The arithmetic circuit is composed of building blocks representing single arithmetic operations over secret shares. Currently, we have implemented addition, multiplication, and comparison operations, as described in~\cite{Cramer_Damgård_Nielsen_2015_Comparison_Protocol}.

The scheduling of the computation is asynchronous and parallelized. Once all inputs for a specific arithmetic operation are ready, the operation is executed. In addition to arithmetic computations such as multiplication, additional tasks are spawned to handle communication-intensive routines, such as homomorphic commitments. As previously mentioned, MPC parties exchange secret shares using the blockchain's P2P network module, while broadcasts are managed by the consensus module.

Due to the high communication complexity associated with secret multiplications, parallel execution of a large number of multiplication gates can saturate network message buffers, as observed in our experiments. To mitigate this, the system enforces a fixed upper bound on the number of parallel multiplication gate evaluations. A FIFO-based queuing mechanism (detailed in the next section) ensures orderly and fair scheduling of multiplication operations as described in the next section.

% Due to the high communication complexity associated with multiplication gates, executing a large number of multiplications in parallel may cause the network message buffer to overflow as we have found. Therefore, the maximum number of parallel multiplication gate executions is limited to a fixed value. To enforce this, a queueing mechanism is to schedule multiplication gate executions in a FIFO order, as described in the next section.

\subsection{MPC Transaction Manager Contract}
A non-payable MPC transaction manager contract called \texttt{MPCTxMgr} that bridges MPC execution with transaction executions by providing the following functionalities:
\begin{enumerate}
    \item Providing an interface for other smart contracts to invoke MPCs
    \item Maintaining the MPC state information for each ongoing MPC transaction
    \item Providing an interface for MPC parties to perform broadcasts and trigger the resumption of MPC transactions when applicable
    \item Coordinating the execution of multiplication gates in the MPC execution module
\end{enumerate}

The first functionality is achieved by \texttt{MPCTxMgr}'s \texttt{enter\_mpc} function, which is called by other smart contracts as illustrated in Section 3. During its execution, \texttt{enter\_mpc} invokes the \texttt{mpc\_start} function of the \texttt{MPCInterface} internal contract, a built-in interface in MPC-EVM's virtual machine that triggers the suspension of an MPC transaction's execution as described in the next section. 

To maintain the MPC state information for ongoing MPC transactions, \texttt{MPCTxMgr} utilizes a storage mapping called \texttt{MPCStates} that maps an MPC transaction's hash to an \texttt{MPCInfo} struct representing an ongoing MPC's state. An \texttt{MPCInfo} struct includes, among others, the following fields:
\begin{enumerate} \renewcommand{\labelenumi}{\alph{enumi})}
    \item A list of MPC parties' addresses
    \item A mapping between an MPC party's id in the address list to the number of parties who sent a transaction accusing it of malicious behaviours when performing an MPC task
    \item A mapping between a multiplication gate ID and the number of parties who broadcasted their readiness for its evaluation and the number who attested its completion, respectively
    \item A mapping between the hash of an MPC result and the number of parties who attested it
\end{enumerate}
A new \texttt{MPCInfo} struct is created and inserted into the mapping when a transaction invokes MPC for the first time, which is determined at the start of \texttt{enter\_mpc} function by checking whether \texttt{MPCStates} contains a corresponding entry for the current transaction's hash. Although the invocation of \texttt{MPCTxMgr}'s \texttt{enter\_mpc} function by a resumed transaction appears to introduce consistency violation risks, this is avoided by \texttt{enter\_mpc}'s control flow. In particular, a resumed transaction only checks for the existence of its \texttt{MPCInfo} struct in \texttt{MPCStates} before directly proceeding to the internal contract function call, thereby avoiding the access of any fields of the \texttt{MPCInfo} struct, which could change between MPC invocations.

Apart from \texttt{enter\_mpc}, \texttt{MPCTxMgr} contains additional functions that are called by MPC parties to perform broadcasts. In addition to fulfilling essential MPC functionalities, such as publicly opening one's share during output reconstruction, broadcasts also facilitate the coordination of MPC multiplication executions in our design, where a fixed number of multiplication gates are executed at a given time to prevent the saturation of network buffers.

When an MPC party \(x\) calls \texttt{MPCTxMgr}'s \texttt{declare\_cheater} function to accuse another party \(y\) of malicious behaviour, the mapping mentioned above as c) aggregates the accusal. Per MPC protocol's design, any MPC party who received accusations from more than \(t\) out of \(n\) parties when performing a particular MPC action is considered malicious. When a party \(x\) sends a broadcast transaction that constitutes the \(t + 1\) accusation to party \(y\), \texttt{MPCTxMgr} will call the internal contract's \texttt{mpc\_finish} function described in the next section, indicating that part \(y\) cheated. Specifically, it returns a result containing dummy values for the MPC outputs followed by two values, with the first one being a 1 that indicates the presence of a malicious party, and the second one being \(y\)'s index in \texttt{MPCInfo}'s array of MPC addresses. This triggers the resumption of the MPC transaction, and the handling of the malicious party is left to the smart contract that invoked the MPC.

In addition to this mapping, \texttt{MPCTxMgr} maintains a global queue of multiplication gates that are ready to be evaluated by the participants of an ongoing MPC. When \(2t+1\) MPC parties have called the \texttt{MPCTxMgr} contract to broadcast their readiness to evaluate a multiplication gate, the gate's information is inserted into the global queue. If the number of ongoing secret multiplications is below the maximum threshold, a newly ready gate is scheduled right away. Otherwise, it is appended to the queue's tail. On the other hand, when \(2t+1\) parties have called \texttt{MPCTxMgr} to attest a gate's completion, a new gate from the queue is scheduled in FIFO order. To instruct the MPC execution module to start an MPC multiplication, \texttt{MPCTxMgr} calls an internal contract function, whose output is an approval message for that gate. This message is forwarded to the MPC execution module from the consensus layer when the broadcast transactions are committed. When a MPC terminates prematurely due to the discovery of a malicious party, its multiplication pending gates in the queue are removed.

If more than \(t\) parties attested to an MPC result \(\text{res}_{MPC}\) via a broadcast transaction, which we call a result attestation transaction, \(\text{res}_{MPC}\) is considered to be verified, and \(\texttt{tx}\) is resumed. Due to the honest majority assumption of the MPC protocol, only the real MPC result will obtain more than \(t\) attestations as there are at most \(t\) malicious parties. Therefore, malicious parties can never make an MPC transaction resume its execution with the wrong MPC results.

Lastly, note that the broadcast functions can only be called by an EOA to prevent arbitrary access to the contract's storage by MPC transactions. This is achieved by checking for the equivalence between \texttt{msg.sender} and \texttt{tx.origin} in each \texttt{MPCTxMgr} function other than \texttt{mpc\_start}.

\subsection{Implementation of Asynchronous MPC Transaction Execution}
\textbf{Internal Contracts}: 
As extending EVM functionalities through the introduction of new opcodes requires ecosystem-wide coordination that involves updating the relevant compilers, IDEs, and SDKs,  Conflux EVM introduces a more practical alternative called \emph{Internal Contracts}.

This concept builds upon Ethereum's precompiled contracts mechanism, which uses natively compiled code to optimize computationally intensive cryptographic operations (e.g., elliptic curve computations) while avoiding the overhead of EVM interpretation. Conflux EVM's Internal Contracts extend this design with additional capabilities, including the direct access of virtual machine internals that enables the implementation of custom virtual machine operations.

Our implementation leverages this by introducing the \texttt{MPCInterface} internal contract deployed at a reserved address, 0x08880...08, replacing the \texttt{enter\_mpc} opcode introduced in Section 4.

For improved developer accessibility, internal contracts are implemented with Solidity ABI compatibility. When invoking an internal contract, the virtual machine bypasses EVM bytecode interpretation. Instead, it passes call parameters and inputs to a predetermined function that adheres to Solidity ABI standards for input parsing and output encoding, enabling seamless integration with standard smart contract interactions. The \texttt{MPCInterface} internal contract exposes a number of Solidity functions as the API interface for MPC-related functionalities, two of which are used for the suspension and resumption of MPC transaction execution, as described below. \\
\textbf{mpc\_start}: the \(\texttt{mpc\_start}(\text{cid}, \text{params})\) function is invoked by \texttt{MPCTxMgr} in its \(\texttt{enter\_mpc}\) function, passing in the arithmetic circuit id \texttt{cid} and public MPC input \texttt{params}. When this function is called, the virtual machine takes different courses of action depending on whether this is the first time that the currently executing transaction \texttt{tx} invokes MPC.

If \texttt{tx} is invoking MPC for the first time, the virtual machine checks that consistency is not violated up to this point in the transaction's execution before its progress is saved. Specifically, the state changes made by \texttt{tx} so far are queried to ensure that no accounts have been modified other than the transaction's sender, the \texttt{MPCTxMgr} contract, and a contract \textit{c} which called \texttt{MPCTxMgr}'s \texttt{enter\_mpc} function. The transaction reverts if this check fails. Otherwise, the virtual machine adds an entry consisting of contract \textit{c}'s address and \texttt{tx}'s hash to a hashmap \(l_C\), which is maintained by the virtual machine to record all currently locked contract accounts. 

Subsequently, state change partitioning is performed. Instead of naively saving all state changes made by the execution of \texttt{tx} up to this point, the virtual machine partitions the state changes into a "pending" set that gets saved and another "ready" set that will be committed with the current block. The "ready" set consists of the addition of a new \texttt{MPCInfo} struct to \texttt{MPCTxMgr}'s mapping \texttt{MPCStates} and the updates to the transaction sender's nonce and balance. If the transaction sender's nonce and balance are not immediately updated, it will not be able to send new transactions before the MPC transaction completes, and this would be disastrous if the sender is also a participant of the invoked MPC. To ensure consistency when accessing account balances, gas refunds are not performed for MPC transactions. It is the transaction senders' responsibility to manage gas provision so that unnecessary losses are minimized. As for the \texttt{MPCTxMgr} contract, the newly created \texttt{MPCInfo} struct is needed by the MPC execution as discussed in the previous section, and therefore this update must also be immediately applied. Furthermore, this will not result in inconsistent accesses afterwards, as the records of MPC executions are intended to remain in \texttt{MPCTxMgr} even if the MPC transaction reverts to provide the blockchain a public view of past MPC executions. 

Finally, the execution context of \texttt{tx} is saved in another hashmap \textit{s} that maps \(\texttt{tx}\)'s hash to the saved context and the number of MPCs it has invoked. When the block containing \texttt{tx} is committed, the consensus layer sends an MPC execution request to the MPCManager, which initiates the MPC execution as described in earlier sections.

On the other hand, a resumed transaction's entry in the hashmap \textit{s} is directly updated with the present state of its execution context. No access constraints are applied at this point, as consistency is continuously enforced during the resumed execution as described in the next section. \\
\textbf{mpc\_finish}: \(\texttt{mpc\_finish}(h_\text{tx}, i, \text{res}_{MPC})\) is called by \texttt{MPCTxMgr} to resume transaction \texttt{tx}'s execution for the \(\textit{i}^{\text{th}}\) time after the result of the \(\textit{i}^{\text{th}}\) MPC has been verified. Note that this call is made from the result attestation transaction \(\texttt{tx}'\) sent by an MPC party that completed the verification of the MPC result, not the saved MPC transaction \texttt{tx}. 

Two cases can occur when \texttt{mpc\_finish} is called. The first case occurs when \textit{i} is equal to the hashmap \(s\)'s record of the number of MPCs invoked by this transaction. In this case, the saved execution context of \texttt{tx} is extracted from \(s\), and \texttt{tx}'s execution resumes by encoding \(\text{res}_{MPC}\) and setting it as the return value expected by the top activation frame on the saved context's callstack. If the execution encounters another call to \(\texttt{mpc\_start}\), the process described above takes place. If the execution completes or reverts, a copy of the modifications made to the locked contract \(c\) is merged with the state changes of \(\texttt{tx'}\), and they will applied to the blockchain when \(\texttt{tx'}\) is committed.

The second case occurs when \textit{i} is less than the hashmap \(s\)'s record of the number of invoked MPCs. This can happen in situations such as the re-execution of the result attestation transactions in another block after the first block that packed them got rejected by the blockchain's consensus. In this case, no resumed execution takes place, and the hashmap \textit{s} is untouched. This is valid because the following three conditions are true: first, an MPC transaction will always be resumed with the correct result of the \(i^{\text{th}}\) MPC due to the honest majority assumption. Second, the locked contract's state when the MPC transaction resumes after the \(i^{\text{th}}\) MPC is always identical to its state when that MPC is invoked. Finally, the resumed execution of \texttt{tx} may not access any accounts other than the locked contract or the \texttt{MPCTxMgr} contract by calling its \texttt{enter\_mpc} function, which does not result in consistency violations as previously discussed in Section 5.3. Because of these three conditions, the resumed execution of \(\texttt{tx}\) after the \(i^{\text{th}}\) MPC is always deterministic and results in the same modifications to the execution context. Therefore, it is safe to directly add the present execution context back to \(s\) when encountering another MPC invocation, as re-executions will not yield a different result. This also implies that it is safe to skip the re-executions.

\subsection{Implementation of Access Control} \label{access_control_impl}
As mentioned earlier, our implementation of MPC-EVM's virtual machine maintains a hashmap \(l_C\) that stores the locked contract's address with the hash of the MPC transaction. In addition to the access constraint applicable for transactions invoking MPC for the first time, MPC-EVM's access control mechanism enforces additional constraints in various scenarios as described below.

When a balance transfer takes place via a transfer transaction or an external message call whose \texttt{msg.value} field is nonzero, the hashmap is queried to ensure that the recipient is not a locked contract. If a contract selfd-sestructs, it may not send its refund to a locked address unless the self-destructing contract is locked and sending the refund balance to itself. 

When the virtual machine executes a \texttt{CALL} or \texttt{BALANCE} opcode, \(l_C\) is first queried to determine if the caller is a locked contract. If it is, \(l_C\) is queried again with the callee address provided by the \texttt{addr} field of the opcode to ensure that it is either the locked contract itself or \texttt{MPCTxMgr} to prevent inconsistent reads or writes to other accounts. If the caller is not a locked contract, the callee address must not be locked for obvious reasons. 

Note that a smart contract is allowed to execute a locked contract's bytecode via the \texttt{CALLCODE} or \texttt{DELEGATECALL} opcode as the execution takes place in the caller's context and only its storage is accessed and/or modified. However, a \texttt{CALLCODE} opcode would still fail if it attempts to execute a locked contract's bytecode while transferring balance to it. This case is covered by the first constraint above.

When the \texttt{CREATE} or \texttt{CREATE2} opcode is encountered, \(l_{C}\) is queried with the deployer account's address. If the address is present, the deployment fails, as allowing a locked contract to deploy a new contract could lead to consistency violations. For instance, if a locked contract transfers balance during contract deployment before invoking MPC, other transactions may read the balance of the address where the new contract will be deployed when the MPC transaction completes, resulting in an inconsistent read.

Finally, note that deadlocks cannot occur in MPC-EVM as a transaction always reverts if it tries to access a locked account; no waiting for lock acquisition takes place.

\section{Evaluations} \label{sec:evaluation}
In this section, we present the empirical evaluation of MPC-EVM, which is driven by the following questions:
\begin{enumerate}
    \item Can MPC-EVM's access control mechanism preserve transaction consistency under various transaction execution scenarios?
    \item What is the impact of executing MPC transactions on the system's throughput?
    \item Are MPC transactions on MPC-EVM robust against malicious behaviours during MPC?
\end{enumerate}
\subsection{Access Control Mechanism}
To validate the correctness of MPC-EVM's access control mechanism, we tested various scenarios in which a transaction attempts to perform operations that violate consistency. To this end, we developed an MPC-invoking smart contract, \texttt{C1}, along with two regular smart contracts, \texttt{C2} and \texttt{C3}, as shown in Listing \ref{lst:locktester}. These contracts feature functions that can be invoked in specific sequences to trigger consistency violation scenarios where the access control mechanism can be tested.
\begin{lstlisting}[caption={Access control test contracts}, label={lst:locktester}]
contract C1 {
    C2 c2;
    MPCTxMgr mpcTxMgr;
    uint public x1;
    address[] mpc_parties;
    uint cid;
    
    constructor (address c2Addr, address[] _mpc_parties, uint circuitId) {
        c2 = C2(c2Addr);
        mpc_parties = _mpc_parties;
        cid = circuitId;
    }

    function callMpc(bool befMPC, bool aftMPC) external {
        x1 = 1;
        if (befMPC) {
            c2.setX2(x1);
        }
        uint[] params = genParams();
        uint[] result = mpcTxMgr.enter_mpc(cid, mpc_parties, params);
        if (aftMPC) {
            c2.setX2(result[0]);
        }
        x1 = 2;
    }

    function setX1(uint val) external {
        x1 = val;
    }
}

contract C2 {
    uint public x2;
    function setX2(uint val) external {
        x2 = val;
    }
}

contract C3 {
    C1 c1;
    uint public x3;
    constructor (address c1Addr) {
        c1 = C1(c1Addr);
    }
    function modifyC1(uint val) external {
        c1.setX1(val);
        x3 = val + 1;
    }
    function getC1Bal() external {
        uint c1Bal_ = address(c1).balance;
        x3 = c1Bal_ + 1;
    }
}
\end{lstlisting}

Contract \texttt{C1} holds a reference to contract \texttt{C2} and provides an MPC-invoking function, \texttt{callMPC()}. This function triggers an MPC at line 20 based on the input circuit ID and, depending on the values of the input parameters \texttt{befMPC} and \texttt{aftMPC}, may call \texttt{C2}'s \texttt{setX2()} function to update its storage variable \texttt{x2} either before or after the MPC execution. Invoking \texttt{C2}'s \texttt{setX2()} function from \texttt{callMPC()} illustrates two consistency violation scenarios: (1) an MPC transaction attempts to modify accounts other than the locked contract or the MPC transaction manager contract before the MPC invocation, and (2) a locked contract attempts to access other accounts. If \texttt{callMPC()} is allowed to execute in this manner, subsequent transactions accessing \texttt{C2} before the MPC transaction completes could read \texttt{x2}'s old value, as the storage modifications made to \texttt{C2} by the MPC transaction would not yet be committed.

In our experiment, callMpc() was invoked with all 4 combinations of the \texttt{befMPC} and \texttt{aftMPC} input parameters. The results are the following: when \texttt{befMPC} is set to true, the transaction reverted before the start of MPC, and when only \texttt{aftMPC} is set to true, the transaction reverted when executing line 21 after the MPC result was obtained. The transaction only succeeded when both inputs parameters are set to false. These are the intended behaviours of MPC-EVM in these scenarios.

On the other hand, contract \texttt{C3}'s \texttt{modifyC1} and \texttt{getC1Bal} function can be called to modify \texttt{C1}'s storage and access its balance, respectively. In another run, an MPC transaction called \texttt{C1}'s \texttt{callMPC} with both input values set to false, and the next transaction called \texttt{C3}'s \texttt{modifyC1} or \texttt{getC1Bal} before the MPC transaction completed, violating consistency by trying to interact with the locked contract \texttt{C1}. In both cases, querying x3's value after the second transaction's execution returned a value of 0, indicating that the state was unchanged due to transaction failure. The locking mechanism's correctness is further validated in the transaction throughput experiments described in the next section where an MPC-invoking function of an MPC contract is constantly called in a loop to start a new MPC. We observed that the transactions calling the MPC contract always fails when an ongoing MPC has already been invoked by the contract.

\subsection{Transaction Throughput}
\textbf{Experiment Setup}: We built a blockchain network consisting of 10 validator nodes running our MPC-EVM implementation. Each validator node runs on an AWS EC2 C6i instance with 16 vCPUS, 32GB of RAM, 128 GB of storage and are connected to each other in a private network.

We developed and deployed five MPC smart contracts: an MPC multiplication contract, an MPC comparison contract, two 10-voter MPC voting contracts, and a 10-bidder MPC auction contract. The voting contract is presented as the motivating example in Section 3, while the multiplication and comparison contracts perform a single multiplication and comparison between two secret shared values, respectively. The MPC auction contract is presented as a case study in the following section. Each smart contract contains a function that invokes an MPC involving 10 parties to perform a specific activity and update the contract storage with the results. The lines of code required to implement the contracts are shown in Table \ref{tab:LoC}. All contracts are written in the standard Solidity language without using any customized syntax.

In addition to the MPC smart contracts, we deployed an ERC-20 token contract and used the Web3 library to pre-generate two streams of non-MPC transactions, one consisting of 2 million Ether transfer transactions and the other one consisting of 2 million ERC-20 token transfer transactions. These transactions are executed to obtain the baseline throughput, and they also execute alongside the MPC transactions when measuring the throughput of the blockchain when executing workloads mixing regular and MPC transactions.

\begin{table}[h]
\centering
\begin{tabular}{|l|c|}
\hline
\textbf{MPC Activity} & \textbf{Line of Code} \\ \hline
Multiplication        & 62                    \\ \hline
Comparison            & 62                    \\ \hline
Voting, 2 Proposals, 10 Voters                & 107                   \\ \hline
Auction, 10 Bidders               & 120                   \\ \hline
\end{tabular}
\caption{Lines of Solidity code in each MPC smart contract}
\label{tab:LoC}
\end{table}

Finally, the JSON-RPC interface of each node is extended with a custom function that reads the pre-generated transactions from a local file and adds them to the node's mempool. This eliminates the delays caused by sending the transactions via the Web3 library, which could prevent the maximum throughput from being reached.
\\
\textbf{Methodology}: The baseline transaction per second (TPS) value is obtained by executing the Ether transfer and ERC-20 token tranfer transactions via a call to the custom JSON-RPC interface function described above. Subsequently, the 10-node network processed workloads consisting of a mixture of regular and MPC transactions. At the start of each run, the MPC smart contracts are deployed, and the MPC parties send transactions to make the required deposit. Afterwards, one of the pre-generated transaction streams is executed while the MPC-invoking function of one MPC contract is repeatedly called in a loop. The call reverts if an MPC transaction is ongoing, and a new MPC will be started immediately if the previous one has completed. This minimizes the amount of time in the experiments where only the regular transactions are executing so that the measured TPS values are as accurate as possible.
In addition to performing the above for each MPC contract, we also experimented with the concurrent execution of multiple MPC transactions by constantly poking the two voting contracts at the same time, resulting in the parallel execution of two voting MPCs. Each run lasted 15 minutes, which was long enough for every type of MPC activity to complete at least once.

% Requires: \usepackage{graphicx}
\begin{table}[h]
    \centering
    \begin{tabular}{|l|c|c|}
    \hline
    & \textbf{ETH Transfer} & \textbf{ERC20 Token Transfer} \\
    \hline
    No MPC & 1905 & 1653 \\
    \hline
    MPC Multiplication & 1903 & 1650 \\
    \hline
    MPC Comparison & 1887 & 1633 \\
    \hline
    1 MPC Voting, 10 Voters & 1884 & 1632 \\
    \hline
    2 MPC Voting, 10 Voters & 1881 & 1613 \\
    \hline
    1 MPC Auction, 10 Bidders & 1860 & 1611 \\
    \hline
    \end{tabular}
    \caption{Transactions per second (TPS) processed by the 10-node network under different workloads. }
    \label{tab:transfer_comparison}
\end{table}

\textbf{Results}: Table 2 presents the TPS achieved by the 10-node network under different workloads. The column represents the type of non-MPC transactions being executed, while the rows represent the MPC workload applied. The baseline throughput is shown in the first row.

Based on the results, workloads that combined MPC transactions with ETH transfers experienced up to a 2.3\% decrease in TPS, whereas workloads combining ERC-20 transfers with MPC transactions experienced a maximum TPS reduction of 2.6\%. It can thus be concluded that the asynchronous processing of MPC transactions coupled with the synchronous execution of regular transactions has minimal impact on the transaction throughput.

\subsection{Case Study: MPC Auction}
The \texttt{MPCAuction} smart contract shown in Listing \ref{lst:auctioncircuit} implements a first-price sealed bid auction on the blockchain where MPC is leveraged to determine the winner and the highest bidding price. Only those two values are publicly revealed at the end of the auction; the bids submitted by other bidders are never revealed, unlike in regular smart contract implementations of sealed bid auctions where all bids are revealed at the end to determine the winner.

When deploying the \texttt{MPCAuction} smart contract, the auction organizer specifies a minimum amount to be deposited by each bidder. In addition to serving as a deterrent against malicious behaviours, a bidder's deposit is also used for payment to the auction organizer if it wins the auction. Similar to the case with the \texttt{MPC\_Vote} contract in Section 3, the bidders are given at least one hour to make the required deposit prior to the start of the auction. When the \texttt{mpcAuction} function is called by the organizer, bidders who failed to make the minimum deposit will have their corresponding entry in the countBids array set to 0 at line 26. Next, \texttt{mpcAuction} calls MPCTxMgr's \texttt{enter\_mpc} function, passing the countBids array as public input to the MPC. This triggers the saving of the current transaction's progress and the invocation of the auction MPC when the current block is committed, as described in earlier sections.

Listing \ref{lst:auctionpseudocode} shows the pseudocode for constructing the arithmetic circuit of an MPC auction involving 10 bidders. The computation takes a secret input array \texttt{x}, which contains the secret-shared bid of each bidder, and another public input array \texttt{w}, which corresponds to the \texttt{countBids} array provided by the auction smart contract. The computation starts by multiplying each secret shared bid \texttt{x[i]} with the corresponding public value \texttt{w[i]}, and it compares each adjacent pair of adjusted bidding prices to determine the higher of the two using the \texttt{MPCCompare} module, as shown in lines 8-12. As \texttt{w[i]} = 0 if the \(i^{\text{th}}\) bidder failed to make the minimum required deposit prior to \texttt{enter\_mpc}'s call, the multiplication effectively eliminate those secret shared bids by setting the underlying value to 0. Subsequently, pairwise comparisons between the output of two previous comparisons are recursively performed as shown in line 15 to 18 until the maximum of the 10 bids is identified.

When the transaction resumes after the MPC results are obtained, post-auction processing is performed in the function \texttt{processAuctionResult}, called at line 33. In addition to checking for the presence of malicious bidders found during the MPC execution, processAuctionResult() also verifies that the highest bidder did not submit a bidding price higher than amount it deposited. If the verification fails, the highest bidder is identified as a cheater, triggering the invocation of the \texttt{processCheater} function at line 46. Otherwise, the payment is deducted from the highest bidder's deposit and can be later withdrawn by the auction organizer. The transaction completes after this point, and the state changes made to the \texttt{MPCAuction} contract are committed to the blockchain with a result attestation transaction as described in Section 5, causing the contract to be unlocked.

\begin{lstlisting}[caption={Simplified Code Snippet of MPC Auction Contract}, label={lst:auctioncircuit}]
contract MPCAuction {
    uint public startableTime;
    MPCTxMgr mpcTxMgr;
    uint public minDeposit;
    address[] bidders;
    uint[] countBid;
    uint cid;
    mapping(address => uint) deposits;
    ...

    constructor(..., uint _minDeposit, uint aucCircId) { 
        startableTime = block.timestamp + 3600;
        cid = aucCircId;
        ...
    }

    function deposit() external payable { .. }
    function withdraw(uint amount) external validAmount { .. }

    function mpcAuction() external isOrganizer validStartTime {
        //bidders who did not make the required deposit have zero weight
        for (uint i = 0; i < bidders.length; ++i) {
            if(deposits[bidders[i]] >= minDeposit) {
                countBid.push(1);
            } else {
                countBid.push(0);
            }
        }
        //invoke MPC auction procedure
        uint[] memory results = mpcTxMgr.enter_mpc(cid, bidders, countBid);
        //process result
        processAuctionResult(results);
    }

    function processAuctionResult(uint[] results) internal {
        if (results[results.length-2] != 0) {
            address cheater = bidders[results[results.length-1]];
            processCheater(cheater);
        }
        else {
            highestBid = results[0];
            uint32 winnerId = results[1];
            highestBidder = bidders[winnerId];
            //highest bidder cheated
            if (highestBid > deposits[highestBidder]) {
                processCheater(highestBidder);
            } else {
                deposits[highestBidder] -= highestBid;
                succeeded = true;
                ...
            }
        }
        ...
    }

    function processCheater(address cheater) internal {
        //distribute the cheater's deposits to other bidders and beneficiary
        ...
    }
    ...
}
\end{lstlisting}

\begin{lstlisting}[caption={Pseudocode for constructing a 10-party auction circuit}, label={lst:auctionpseudocode}]
/* x[i] is the ith bidder's bid value, secret shared
   w[i] is a 0/1 value, indicating if it should be counted */
fn 10_party_auction(x[10; 1]: Secret, w[10; 1]: Public) -> max_bid, winner {
    max_val = new Secret[9]
    max_id = new Secret[9]
	
    //Parallel pairwise comparison between inputs bids  
    max_val[0], max_id[0] = MPCCompare(MPCMultByConst(x[0][0], w[0]), 0, MPCMultByConst(x[1][0], w[1]), 1)
    max_val[1], max_id[1] = MPCCompare(MPCMultByConst(x[2][0], w[2]), 2, MPCMultByConst(x[3][0], w[3]), 3)
    max_val[2], max_id[2] = MPCCompare(MPCMultByConst(x[4][0], w[4]), 4, MPCMultByConst(x[5][0], w[5]), 5)
    max_val[3], max_id[3] = MPCCompare(MPCMultByConst(x[6][0], w[6]), 6, MPCMultByConst(x[7][0], w[7]), 7)	
    max_val[4], max_id[4] = MPCCompare(MPCMultByConst(x[8][0], w[8]), 8, MPCMultByConst(x[9][0], w[9]), 9)
	
    //Comparison between intermediate results
    max_val[5], max_id[5] = MPCCompare(max_val[0], max_id[0], max_val[1], max_id[1])
    max_val[6], max_id[6] = MPCCompare(max_val[2], max_id[2], max_val[3], max_id[3])
    max_val[7], max_id[7] = MPCCompare(max_val[4], max_id[4], max_val[5], max_id[5])
    max_val[8], max_id[8] = MPCCompare(max_val[6], max_id[6], max_val[7], max_id[7])

    max_bid = max_val[8]
    winner = max_id[8]
}
\end{lstlisting}

\subsection{Handling Malicious MPC Parties}
Lastly, we verified the robustness of MPC-EVM against malicious MPC parties. We designed an experiment where one node's MPC execution module was modified so that it distributes inconsistent polynomials to the other parties when performing a commitment to a secret using the protocol described in \cite{Cramer_Damgård_Nielsen_2015_IT_Protocol}. This simulated a malicious party who attempted to sabotage the MPC protocol. Investigating the execution trace on each node reveals that the other parties correctly noticed the inconsistency and sent transactions disputing the point where the value they evaluated did not match that of the others. When the malicious party publicly opened the polynomials it shared in response to the disputes, the honest party correctly sent broadcast transactions to \texttt{MPCTxMgr} accusing the malicious party of dishonest behaviour. The transaction was resumed at this point as a quorum of accusations was collected. Accessing the MPC's result, which was stored by the test contract, showed the index corresponding to the malicious party's index in the list of MPC party addresses.
\section{Related Work} \label{sec:related_work}
Prior studies have investigated various ways of integrating MPC with blockchains. Some early works, such as Enigma \cite{shrobe2018enigma}, utilized blockchains as secure broadcast channels for performing multiparty computations, while others such as \cite{bentov2014use} and \cite{kiayias2016fair} leveraged the monetary incentives offered by cryptocurrencies to encourage honest behaviours in MPC protocols. 

More recent works have explored the construction of privacy-preserving smart contracts using MPC. Among those, the Ratel \cite{li2024ratel} framework is the most relevant to this work as it aims to realize asynchronous execution of MPC by smart contracts. It extended the Solidity language with MPC-enabling bytecodes and developed the Ratel language. Programs written in Ratel get compiled into the bytecode of an MPC-invoking smart contract to be deployed as well as an MP-SPDZ \cite{cryptoeprint:2020/521} library program. The MPC library program is called to perform the MPC when triggered by the on-chain execution. Additionally, it enables the MPC parties to store private states offchain in a database, and it developed a crash recovery mechanism that allows crashed parties to rejoin the MPC. However, the system does not prevent other transactions from accessing and modifying the MPC-invoking smart contract during the MPC execution, a problem which is addressed in our design. 

Another related work is zkHawk \cite{banerjee2021zkhawk}, which builds on top of Hawk \cite{kosba2016hawk} and implements private smart contracts as off-chain MPC coupled with on-chain commitment and proof verifications. In this framework, users commit to their input output balances via homomorphic encryption, and execute the smart contract's logic off-chain via MPC. Upon MPC's completion, users collaboratively generate proofs of the execution's correctness and the equality of the input/output balances and upload them to the smart contracts for verification. Compared with our work, zkHawk encapsulates the entire smart contract logic in an offline MPC and does not support the interleaving of MPC with smart contract execution.

Additional works have investigated the design of novel MPC protocols that can be implemented in a blockchain system to preserve input privacy. Eagle \cite{baum2023eagle} developed an UC-secure protocol building on top of Insured MPC \cite{baum2020insured} that can keep both the user's input data and balance private while executing the contract logic via MPC. Gage MPC \cite{almashaqbeh2021gage} designed a garbled-circuit based non-interactive MPC (NIMPC) protocol and implemented an Ethereum auction based on this protocol. However, strong privacy is only guaranteed in the short term in this model.

In addition to MPC, other cryptographic techniques, such as fully homomorphic encryption (FHE), have also been applied to achieve input privacy on blockchain activities that require input from multiple parties. In particular, \cite{cryptoeprint:2022/1119} implemented sealed-bid auction using threshold FHE, where computation is performed directly on the encrypted inputs and only the auction winner is collaboratively opened by the bidders. However, the study acknowledged that the blockchain state variables could be arbitrarily modified by other transactions during the FHE execution.
\section{Conclusion} \label{sec:conclusion}
We presented {\name}, the first blockchain platform that natively supports secure multiparty computations (MPC) within smart contracts. {\name} introduces a novel asynchronous execution model for MPC-enabled transactions, pausing their execution at MPC invocation points, saving intermediate execution states, and seamlessly resuming transactions upon MPC completion. Additionally, {\name} maintains the consistency guarantees inherent to the Ethereum Virtual Machine (EVM) through a novel access control policy, ensuring consistent execution for contracts involved in ongoing MPC computations.

Our evaluation demonstrates that {\name} is sufficiently expressive to support practical use cases such as decentralized voting and auctions. By asynchronously executing high-latency MPC computations, {\name} achieves robust performance, experiencing less than a 3\% reduction in transaction throughput in a 10-node setting when MPC-enabled transactions run concurrently with regular transactions. These promising results highlight the potential for deeper integration of MPC into blockchain systems, paving the way for the development of privacy-preserving decentralized applications.

%\section{Data Availability}

%We will make the source code of {\name} and all experimental data publicly available as part of this paper's artifact. Upon %acceptance, we intend to submit the artifact %for evaluation.

\bibliographystyle{ACM-Reference-Format} % Specify the ACM reference style
%\bibliography{references}
%%% -*-BibTeX-*-
%%% Do NOT edit. File created by BibTeX with style
%%% ACM-Reference-Format-Journals [18-Jan-2012].

\end{document}